\documentstyle{article}
\def\npag{\thanks{
This paper was done under the auspices of a CNCSIS Grant.}}
\date{\npag}

\title{ Deformation of the Planetary Orbits \\ 
Caused by the Time Dependent Gravitational \\
Potential in the Universe}

\author{Kostadin Tren\v{c}evski$^*$
\\ Faculty of Natural Sciences and Mathematics,\\ 
P.O.Box 162, 1000 Skopje, Macedonia} 

\begin{document}
\maketitle

PACS number: 98.80.Jk
{\renewcommand{\thefootnote}{}%
\footnote{$^*$Electronic address: kostatre@iunona.pmf.ukim.edu.mk}%
\setcounter{footnote}{0}}%

\begin{abstract} 
In the recent paper \cite{KT1}, assuming 
a linear change of a gravitational 
potential $V$ in the universe, i.e. $\Delta V=-c^2H\Delta t$, 
are explained both the Hubble red shift and the anomalous 
acceleration $a_P$ from the spacecraft Pioneer 10 and 11 
\cite{JDA}. The change of the potential $V$ causes an accelerated time 
which is easily seen by the Hubble red shift. But the change of the potential
causes also change of the distances in the galaxy, and hence modification 
on the planetary orbits. In this paper 
it is shown that the planetary orbits are not axially symmetric, 
neglecting the relativistic corrections. The angle from the perihelion to 
the aphelion is $\pi-\frac{H\Theta}{3e\pi}$, while the angle from the 
aphelion to the perihelion is $\pi+\frac{H\Theta}{3e\pi}$,
where $\Theta$ is the orbit period, $e$ is the eccentricity 
of the elliptic trajectory and 
$H$ is the Hubble constant. There is no perihelion precession 
caused by the time dependent gravitational potential $V$. 
The quotient $\Theta_2:\Theta_1$ of two consecutive orbit periods 
$\Theta_1$ and $\Theta_2$ 
is equal to 
$\Theta_2:\Theta_1=1+\frac{\Theta H}{3}.$
This formula is tested for the orbit period of the pulsars 
B1534+12 and 1855+09 which have very good timing, and the results are 
satisfactory. 
%
\end{abstract}

In the recent paper \cite{KT1} is given an assumption that 
in the universe there is a gravitational potential $V$, which 
decreases linearly (or almost linearly), 
such that $\Delta V=-c^2H\Delta t$, where $H$ is the 
Hubble constant. This change of the potential just causes the Hubble 
red shift $\nu =\nu_0 \Bigl (1-\frac{RH}{c}\Bigr )$ and the large 
velocities among the galaxies are apparent. 
The change of the potential is probably caused by the 
dark energy in the universe, which is about $67\%$ \cite{F}. 

Note that the gravitational potential is assumed to be larger near 
the gravitational bodies, than the potential where there is no gravitation. 
For example, near the spherical body we accept 
that the potential is $\frac{GM}{r}$. If we accept that this potential is 
$-\frac{GM}{r}$, then for the time dependent potential it should be 
$\Delta V=c^2H\Delta t$. 

The change of the gravitational potential causes nonuniform space-time. 
Let us denote by $X,Y,Z,T$ our natural coordinate system in the 
space-time deformed by the gravitational potential $V$ and 
let us denote by $x,y,z,t$ the normed coordinates of an imagine 
coordinate system, where the space-time is "uniform". 
This physically occurs for example, if we travel with a lift 
away from the ground with slow velocity $v=2,44$ cm/s. Then we have 
the same time dependent change of the gravitational potential 
like in the universe, but 
in the lift we have also a space dependent gravitational potential, 
which causes the acceleration toward the Earth. 

According to the general relativity we have the following equalities 
\begin{equation}
dx=\Bigl (1+\frac{V}{c^2}\Bigr )^{-1}dX = \Bigl (1+tH\Bigr )dX, \label{1}
\end{equation}
\begin{equation}
dy=\Bigl (1+\frac{V}{c^2}\Bigr )^{-1}dY = \Bigl (1+tH\Bigr )dY,\label{2}
\end{equation}
\begin{equation}
dz=\Bigl (1+\frac{V}{c^2}\Bigr )^{-1}dZ = \Bigl (1+tH\Bigr )dZ,\label{3}
\end{equation}
\begin{equation}
dt=\Bigl (1-\frac{V}{c^2}\Bigr )^{-1}dT = \Bigl (1-tH\Bigr )dT.\label{4}
\end{equation} 
Hence we obtain 
$$
\Bigl (\frac{dX}{dT},\frac{dY}{dT},\frac{dZ}{dT}\Bigr )=
\Bigl (\frac{dx}{dt},\frac{dy}{dt},\frac{dz}{dt}\Bigr )
(1-2tH)
$$
and by differentiating this equality we get 
$$
\Bigl (\frac{d^2X}{dT^2},\frac{d^2Y}{dT^2},\frac{d^2Z}{dT^2}\Bigr )=$$
\begin{equation}
=\Bigl (\frac{d^2x}{dt^2},\frac{d^2y}{dt^2},\frac{d^2z}{dt^2}\Bigr )
-3HT\Bigl (\frac{d^2X}{dT^2},\frac{d^2Y}{dT^2},\frac{d^2Z}{dT^2}\Bigr )
-2H\Bigl (\frac{dX}{dT},\frac{dY}{dT},\frac{dZ}{dT}\Bigr ), \label{a} 
\end{equation}
where 
$\Bigl (\frac{d^2x}{dt^2},\frac{d^2y}{dt^2},\frac{d^2z}{dt^2}\Bigr )$ 
is the Newtonian acceleration. 
In normed coordinates $x,y,z,t$ there is no any acceleration caused by the 
time dependent gravitational potential, because the gradient of $V$ 
vanishes. Thus, according to the 
coordinates $X,Y,Z,T$ appears an additional acceleration 
\begin{equation}
-3HT\Bigl (\frac{d^2X}{dT^2},\frac{d^2Y}{dT^2},\frac{d^2Z}{dT^2}\Bigr )
-2\Bigl (H\frac{dX}{dT},H\frac{dY}{dT},H\frac{dZ}{dT}\Bigr ) .\label{5}
\end{equation}
Using the acceleration (\ref{5}) in \cite{KT1} is explained the 
anomalous acceleration $a_P\approx Hc$, which causes an anomalous 
frequency shift of the radio signals which are sent to the spacecraft 
Pioneer 10 and 11 and re-transmitted to the Earth \cite{JDA}. 

The acceleration (\ref{5}) appears also in the planetary orbits. 
In order to study that, first we shall consider two special cases of 
deformations, like basic generators of deformations. 
Further we shall neglect the relativistic deformations
(for example relativistic perihelion precession), in order to emphasize the 
deformations caused by the time dependent gravitational potential. We shall 
consider three parameters of the planetary orbits: 1. the angle 
between the radii given by the perihelion and aphelion, 
i.e. its departure $\Delta \varphi$
from $\pi$; 
2. the perihelion precession; and 3. 
the quotient $\Theta_2:\Theta_1$, where $\Theta_1$ and $\Theta_2$ 
are two consecutive orbit periods of the planet. 

I. Assume that instead of (1), (2), (3) and (4) we have the following 
coordinate transformations 
\begin{equation} 
x=(1+\lambda tH)X,\; y=(1+\lambda tH)Y, 
\; z=(1+\lambda tH)Z, \; t=T,\; (\lambda =const.),\label{6} 
\end{equation}
which means that we have only change in the space coordinates, 
and there is no change in the time coordinate. 
According to the normed coordinates
$x,y,z,t$ the trajectory in the plane of motion is given by 
\begin{equation}
\frac{1}{r}=\frac{\rho_1+\rho_2}{2}+\frac{\rho_1-\rho_2}{2}\cos \varphi ,
\label{7} 
\end{equation} 
where $\rho_1=1/r_1$ and $\rho_2=1/r_2$ are constants, 
just like in the Newtonian theory. 

According to (\ref{6}), the equality (\ref{7}) becomes 
\begin{equation} 
\frac{1}{R} = (1+\lambda TH)
\Bigl [\frac{\rho_1+\rho_2}{2}+
\frac{\rho_1-\rho_2}{2}\cos \varphi \Bigr ].\label{8} 
\end{equation} 
The angle $\varphi$ is a common parameter for both coordinate systems. 
The equation 
$\frac{d\rho}{d\varphi}=0$ for the extreme values of $1/R$, 
according to (\ref{8}), yields to 
$$\frac{d}{d\varphi}\Bigl [
\frac{\rho_1+\rho_2}{2}(1+\lambda tH)+
\frac{\rho_1-\rho_2}{2}(1+\lambda tH)\cos \varphi \Bigr ]
=0,$$
$$-\frac{\rho_1-\rho_2}{2}(1+\lambda tH)\sin \varphi +
\Bigl [\frac{\rho_1+\rho_2}{2}+\frac{\rho_1-\rho_2}{2}\cos \varphi \Bigr ]
\frac{dt}{d\varphi}\lambda H=0,$$
$$\sin \varphi = \Bigl [\frac{\rho_1+\rho_2}{\rho_1-\rho_2}+\cos \varphi 
\Bigr ]\frac{\lambda H}{\frac{d\varphi}{dt}}.$$
The solution of this equation of $\varphi $ when $\varphi \approx 0$ 
can be obtained approximately 
by putting $\cos \varphi =1$ and hence 
\begin{equation}
\varphi_1=\frac{2\rho_1}{\rho_1-\rho_2}
\frac{\lambda H}{(\frac{d\varphi}{dt})_{per.}}=
\frac{2r_1\lambda H}{(\rho_1-\rho_2)C},\label{9}
\end{equation}
where $C=r^2\frac{d\varphi}{dt}=const.$ according to the second Kepler's 
law. The solution of $\varphi$ when $\varphi\approx \pi$ can be obtained 
approximately by putting $\cos \varphi =-1$, i.e. 
\begin{equation}
\varphi_2=\pi - \frac{2\rho_2}{\rho_1-\rho_2}
\frac{\lambda H}{(\frac{d\varphi}{dt})_{aph.}}
=\pi -\frac{2r_2\lambda H}{(\rho_1-\rho_2)C}.\label{10}
\end{equation}

Hence 
$$\varphi_2-\varphi_1=\pi -\frac{2(r_2+r_1)\lambda H}
{\frac{r_2-r_1}{r_1r_2}C}=\pi -\frac{2r_1r_2\lambda H}{Ce},$$
where $e$ is the eccentricity. Assuming that the eccentricity of the ellipse 
is small, then 
$\frac{C}{r_1r_2}\approx \frac{d\varphi}{dt}\approx \frac{2\pi}{\Theta}$, 
where $\Theta$ is the orbit period of the planet, and hence 
\begin{equation}
\Delta \varphi =\varphi _2-\varphi _1- \pi\approx - 
\frac{\lambda H\Theta}{\pi e}.\label{11}
\end{equation}

Further, for the precession of the perihelion we obtain the angle 
\begin{equation}
\frac{2\rho_1}{\rho_1-\rho_2}\Bigl [
\frac{\lambda H}{(\frac{d\varphi}{dt})_{\Theta}}-
\frac{\lambda H}{(\frac{d\varphi}{dt})_{0}}\Bigr ]=0, 
\end{equation}
neglecting the terms of order $H^2$. 
Hence the angle from the aphelion to the perihelion is equal to 
\begin{equation}
\pi +\frac{2r_1r_2\lambda H}{Ce}\approx 
\pi + \frac{\lambda H\Theta}{\pi e}, 
\end{equation} 
and the planetary orbit is not axially symmetric. 

Note that in the previous considerations it was assumed 
that $H$ does not change with the time. If $H$ changes with the time,
then will appear a slight departure of order $H^2$ and it is negligible. 

Since $t=T$ in this case, it is 
\begin{equation}
\Theta_2:\Theta_1=1.\label{13}
\end{equation} 

It is easy to see that the accelerations in both 
coordinate systems are related by 
\begin{equation}
\Bigl (\frac{d^2X}{dT^2},\frac{d^2Y}{dT^2},\frac{d^2Z}{dT^2}\Bigr ) 
=(1-\lambda TH)
\Bigl (\frac{d^2x}{dt^2},\frac{d^2y}{dy^2}\frac{d^2z}{dt^2}\Bigr )
-2\lambda H\Bigl (\frac{dX}{dT},\frac{dY}{dT},\frac{dZ}{dT}\Bigr ).\label{14}
\end{equation}

II. Assume that instead of (1), (2), (3), and (4) we have the following 
coordinate transformations 
\begin{equation} 
x=X,\quad y=Y, \quad z=Z, \quad dt=(1-\mu tH)dT,
\quad (\mu =const.),\label{15}
\end{equation}
which means that we have only change in the time coordinate. Thus, the 
planetary orbit in the $(R,\varphi )$-plane is an ellipse like 
according to the $x,y,z,t$ coordinate system. 

Using that $T=t+\frac{\mu}{2}Ht^2$, for the quotient 
$\Theta_2:\Theta_1$ which corresponds to $t_2=t_1$ 
for the orbit period we obtain 
$$\Theta_2:\Theta_1=\Bigl ([2\Theta +\frac{\mu}{2}H(2\Theta )^2]-
[\Theta +\frac{\mu}{2}H\Theta^2]\Bigr ):
\Bigl (\Theta +\frac{\mu}{2}H\Theta^2\Bigr )
=1+\mu \Theta H,$$
i.e. 
\begin{equation} 
\Theta_2:\Theta_1=1+\mu \Theta H.\label{16} 
\end{equation} 

Finally, note that the equations (\ref{15}) imply that 
$$\frac{dX}{dT}=\frac{dx}{(1+\mu Ht)dt} = \frac{dx}{dt}(1-\mu tH),$$
$$\frac{d^2X}{dT^2}=\frac{d}{(1+\mu Ht)dt}
\Bigl (\frac{dx}{dt}(1-\mu tH)\Bigr )
=(1-\mu tH)^2\frac{d^2x}{dt^2}-\mu H\frac{dX}{dT},$$
and analogously is true for $Y$ and $Z$ coordinates. Indeed, 
\begin{equation}
\Bigl (\frac{d^2X}{dT^2},\frac{d^2Y}{dT^2},\frac{d^2Z}{dT^2}\Bigr ) = 
(1-2\mu TH)\Bigl (\frac{d^2x}{dt^2},\frac{d^2y}{dt^2},\frac{d^2z}{dt^2}
\Bigr )-\mu H\Bigl (\frac{dX}{dT},
\frac{dY}{dT},\frac{dZ}{dT}\Bigr ).\label{17}
\end{equation}

Combining both special cases I. and II. we obtain the following conclusion: 
{\em If the acceleration is given by 
$$\Bigl (\frac{d^2X}{dT^2},\frac{d^2Y}{dT^2},\frac{d^2Z}{dT^2}\Bigr ) = $$
\begin{equation} 
=(1-(\lambda +2\mu )TH) 
\Bigl (\frac{d^2x}{dt^2},\frac{d^2y}{dt^2},\frac{d^2z}{dt^2}\Bigr )-
(2\lambda +\mu )H 
\Bigl (\frac{dX}{dT},\frac{dY}{dT},\frac{dZ}{dT}\Bigr ),\label{18}
\end{equation}
then the angle $\Delta \varphi$ is given by (\ref{11}), 
there is no precession of the perihelion and the quotient 
$\Theta_2:\Theta_1$ is given by} (\ref{16}). 


Let us consider the space-time in the universe, where 
we accepted the equalities (\ref{1}), (\ref{2}), (\ref{3}) and (\ref{4}). 
If we put $\lambda =\frac{1}{3}$ and $\mu =\frac{4}{3}$, then 
the equation (\ref{18}) becomes just the equation (\ref{a}). 
Thus, for the parameter $\Delta \varphi$ we obtain 
\begin{equation} 
\Delta \varphi = -\frac{H\Theta}{3\pi e}.\label{19} 
\end{equation} 

According to the formula (\ref{19}) the angle $\Delta \varphi$ for Mercury 
is equal to $-9\times 10^{-12}$ radians. For a comparison, the perihelion 
precession according to the general relativity is about $4,8\times 10^{-7}$
radians. {From} the formula (\ref{19}) 
we see that the angle $\Delta \varphi$ can easier be measured 
if we consider a planet with small eccentricity and long period $\Theta$. 
For example, if we consider the orbit of the Earth and use that 
$e\approx 1/60$, and for $H$ take the value 
$(14\times 10^9$ years$)^{-1}$, then we 
obtain that $\Delta \varphi \approx -0,45\times 10^{-9}$ radians, i.e. 
about $-(10^{-4})''$. Hence the aphelion is displaced for 
1AU$\times \Delta \varphi\approx -67$m. The value of 
$\Delta \varphi $ for Neptune is about 300 
times larger than the value for the Earth and can easily be measured, 
but the half of the orbit period is 83 years and it is too long. 
The value of $\Delta \varphi $ for Venus is 1,5 times larger than the 
value for the Earth. 

According to (\ref{aa}), the quotient $\Theta_2:\Theta_1$ of two consecutive 
orbit periods is $1+\mu \Theta H=1+\frac{4}{3}\Theta H$,
and it should be "normed" with 
$1+\Theta H$ according to (\ref{4}), because in the coordinate system 
$X,Y,Z,T$ after each orbit period the time is faster $1+\Theta H$ times. 
Hence the required quotient is equal to 
\begin{equation} 
{\Theta _2:\Theta_1}=\frac{1+\frac{4}{3}\Theta H}{1+\Theta H}
=1+\frac{1}{3}\Theta H.\label{20}
\end{equation} 
If we consider the planet Earth, i.e. $\Theta $ is one year, then 
$$\Theta _2-\Theta _1 =\frac{1}{3}\Theta^2H\approx 0,001 \; \hbox{s}.$$
The difference $\Theta_2-\Theta_1$ for Pluto is about one minute and it is 
easy to be detected, but two orbit periods is too long time. 

Formula (\ref{20}) can be applied for double stars, for example the 
binary pulsars. {From} (\ref{20}) we get 
\begin{equation}
\dot{P_b} = \frac{1}{3}PH,\label{a21}
\end{equation}
where $P_b=\Theta$ is the orbit period of the binary pulsar. While the 
gravitational radiation decreases the orbit period, 
(\ref{a21}) increases. This increment very often 
is much smaller than the decay caused by the gravitational radiation. 
For example, for the Hulse-Taylor binary pulsar PSR B1913+16 \cite{WT}
the increment is $1\%$ of the decay of the orbit period and it is 
difficult to be detected. So we choose a binary pulsar with longer 
orbit period, and/or small decay caused by the gravitational 
radiation. Such examples are the pulsars PSR B1534+12 \cite{S} 
and PSR B1885+09 \cite{KTR}. Moreover, 
both binaries have very good timings and hence are convenient for 
precise tests. In case of PSR B1534+12 
the decay of $\dot{P_b}$ caused by the gravitational radiation 
is about $-0,1924\times 10^{-12}$, while the increment caused by 
(\ref{a21}) is about $0,027\times 10^{-12}$, and hence together we have 
decay about $-0,1654\times 10^{-12}$. On the other side, the measured 
value of $\dot{P_b}$ is $(-0,174\pm 0,011)\times 10^{-12}$, 
where the galactic corrections are included. 
Hence there is agreement of both results. The agreement is much better 
for the pulsar PSR B1885+09 \cite{KTR}. In this case it is found that 
(formulae (18) and (20) in \cite{KTR}) 
\begin{equation}
-\frac{1}{2P_b}(\dot{P}_b^{obs}-\dot{P}_b^{exp})=
(-9\pm 18)\times 10^{-12}\frac{1}{\hbox {yr}}, \label{a22}
\end{equation}
where $\dot{P}_b^{obs}$ and $\dot{P}_b^{exp}$ are the observed and 
expected values respectively. 
According to (\ref{a21}), $\dot{P}_b^{obs}-\dot{P}_b^{exp}=P_bH/3$, and the 
left side of (\ref{a22}) becomes 
$-11,9\times 10^{-12}\frac{1}{\hbox {yr}}$. 
Note that here it is taken $\dot{P}_b^{exp}=0$, because the influence 
from the gravitational radiation is negligible. Moreover, the kinematic 
galactic corrections are one order smaller than (\ref{a22}). Thus, in case 
of PSR B1885+09 the significant part of the measured value of 
$\dot{P}_b$ is caused by the formula (\ref{a21}). This example well confirms
the presented theory. 



\end{document}